\documentclass{PoS}

\title{Study of H-dibaryon mass in Lattice QCD}

\ShortTitle{Study of $H$-dibaryon mass in Lattice QCD}

\author{\speaker{Takashi Inoue}\\
        Nihon University, College of Bioresource Sciences, Kanagawa 252-0880, Japan\\ 
        E-mail: \email{inoue.takashi@nihon-u.ac.jp}}

\author{for HAL QCD Collaboration}

\abstract{
After a brief review of discovery of the $H$-dibaryon in lattice QCD, 
effect of the flavor $SU(3)$ symmetry breaking on the $H$-dibaryon is studied 
by basing on the baryon-baryon ($B\!B$) interactions extracted from QCD on the lattice.
The Schr\"{o}dinger equation for $\Lambda\Lambda$-$N\Xi$-$\Sigma\Sigma$ coupled-channel 
is solved with the physical baryon masses and the potentials obtained from QCD at the flavor $SU(3)$ limit.
A resonant $H$-dibaryon is found between $\Lambda\Lambda$ and $N\Xi$ thresholds in this treatment.  
}

\FullConference{The 30 International Symposium on Lattice Field Theory - Lattice 2012,\\
		June 24-29, 2012\\
		Cairns, Australia}

\begin{document}

\section{Introduction}

 The $H$-dibaryon, which has baryon number $B=2$ and strangeness $S=-2$, 
 predicted by R.~L.~Jaffe in 35 years ago~\cite{Jaffe:1976yi},
 is one of the most famous candidates of exotic-hadron. 
 This prediction was based on two observations:
 (i) that the quark Pauli exclusion can be completely avoided in the flavor-singlet ($uuddss$) six quark,
 and (ii) that the one-gluon-exchange interaction produce a large attraction for the flavor-singlet
 six quark~\cite{Jaffe:1976yi,Sakai:1999qm}.
 Search for the $H$-dibaryon is one of the most important challenge
 for theoretical and experimental physics of the strong interaction and the quantum chromodynamics (QCD).
 At present, it is not clear whether there exists the $H$-dibaryon in nature or not.
 Although deeply bound $H$-dibaryon with the binding energy $B_H > 7 $ MeV from the $\Lambda\Lambda$ threshold
 has been ruled out by the discovery of the double $\Lambda$ hypernuclei,
 $_{\Lambda \Lambda}^{\ \ 6}$He~\cite{Takahashi:2001nm}, 
 possibilities of a shallow bound state or a resonance state in this channel still remain~\cite{Yoon:2007aq}.

 One promising approach to clarify the $H$-dibaryon is a numerical simulation of QCD on the lattice.
 There was several lattice QCD calculations on the $H$-dibaryon as reviewed in ref.~\cite{Wetzorke:2002mx}
 (see also recent works~\cite{Luo:2007zzb,Beane:2010hg,Beane:2011iw,Beane:2012vq}).
 However, there was a serious problem in studying dibaryons on the lattice:
 To accommodate two baryons inside the lattice volume,
 the spatial lattice size $L$ should be large enough. 
 Once  $L$ becomes large, however, energy levels of two baryons become dense,
 so that isolation of the ground state from the excited states is very difficult
 (quite a large imaginary-time $t$ is required).
 All the previous works on dibaryons more or less face this issue.
 While, we employed our original breakthrough approach and searched the $H$-dibaryon,
 and then found a bound $H$-dibaryon in the flavor $SU(3)$ limit~\cite{Inoue:2010es,Inoue:2011ai}.
 Because the flavor $SU(3)$ breaking complicate calculation, we began with the flavor $SU(3)$ limit.
 We review our lattice QCD results briefly in the next section. 
 In section 3, we estimate effect of the flavor $SU(3)$ breaking on the $H$-dibaryon.

\section{H-dibaryon from full QCD simulations on the lattice}

\begin{table}[t]
\caption{Lattice QCD parameters: number of sites, the inverse coupling $\beta$,
the clover coefficient $c_{\rm sw}$, the lattice spacing $a$ and the spatial lattice extent $L$.}
\label{tbl:lattice}
\bigskip
\centering
 \begin{tabular}{c|c|c|c|c}
   \hline
    sites             & ~~ $\beta$ ~~ & ~~ $c_{\rm sw}$ ~~ & ~ $a$ [fm] ~ & ~$L$ [fm]~  \\
   \hline 
    $32^3 \times 32$  &   1.83  &   1.761   &  0.121(2)  & 3.87  \\
   \hline
 \end{tabular}
\bigskip
\newline
\end{table}

 As shown in refs.~\cite{Ishii:2006ec,Aoki:2009ji,HALQCD:2012aa},
 one can define and extract potential of two-hadron interaction from QCD
 by using the Nambu-Bethe-Salpeter (NBS) wave function measured on the lattice.
 For example, for a single S-wave channel case, the leading order of the derivative expansion
 of the potential can be obtained as
\begin{equation}
  V(\vec r) = \frac{1}{2\mu}\frac{\nabla^2 \Psi(\vec r, t)}{\Psi(\vec r, t)} - 
              \frac{\frac{\partial}{\partial t} \Psi(\vec r, t)}{\Psi(\vec r, t)} - M_1 - M_2 
\label{eqn:vr}
\end{equation}
 where $\Psi(\vec r, t)$ is a measured NBS wave function, 
 and $M_1$, $M_2$ and $\mu$ are measured hadron masses and the reduced mass of them respectively.

 For lattice QCD simulations with dynamical quarks in the flavor $SU(3)$ limit,
 we have generated gauge configurations at five different values of quark masses.
 We have employed the renormalization group improved Iwasaki gauge action
 and the non-perturbatively $O(a)$ improved Wilson quark action. 
 Our simulation parameters are summarized in Table~\ref{tbl:lattice}.
 Hadron masses measured on each ensemble, together with the quark hopping parameter $\kappa_{uds}$,
 are given in  Table~\ref{tbl:mass}.

\begin{table}[t]
\caption{Quark hopping parameter $\kappa_{uds}$ and measured hadron masses:
$M_{\rm ps}$ for pseudo-scalar meson octet, $M_{\rm vec}$ for vector meson octet, 
$M_{\rm bar}$ for baryon octet.}
\label{tbl:mass}
\bigskip
\centering
 \begin{tabular}{c|c|c|c|c}
   \hline
    $\kappa_{uds}$  & ~$M_{\rm ps}$ [MeV]~ &  ~$M_{\rm vec}$ [MeV]~& ~ $M_{\rm bar}$ [MeV]~ &
   ~$N_{\rm cfg}\,/\,N_{\rm traj}$~ \\
   \hline 
     ~0.13660~ &   1170.9(7) &   1510.4(0.9) & 2274(2) & 420\,/\,4200 \\
     ~0.13710~ &   1015.2(6) &   1360.6(1.1) & 2031(2) & 360\,/\,3600 \\
     ~0.13760~ & ~\,836.5(5) &   1188.9(0.9) & 1749(1) & 480\,/\,4800 \\
     ~0.13800~ & ~\,672.3(6) &   1027.6(1.0) & 1484(2) & 360\,/\,3600 \\
     ~0.13840~ & ~\,468.6(7) & ~\,829.2(1.5) & 1161(2) & 720\,/\,3600 \\
   \hline
 \end{tabular}
\bigskip
\newline
\end{table}

\begin{figure}[t]
\includegraphics[width=0.475\textwidth]{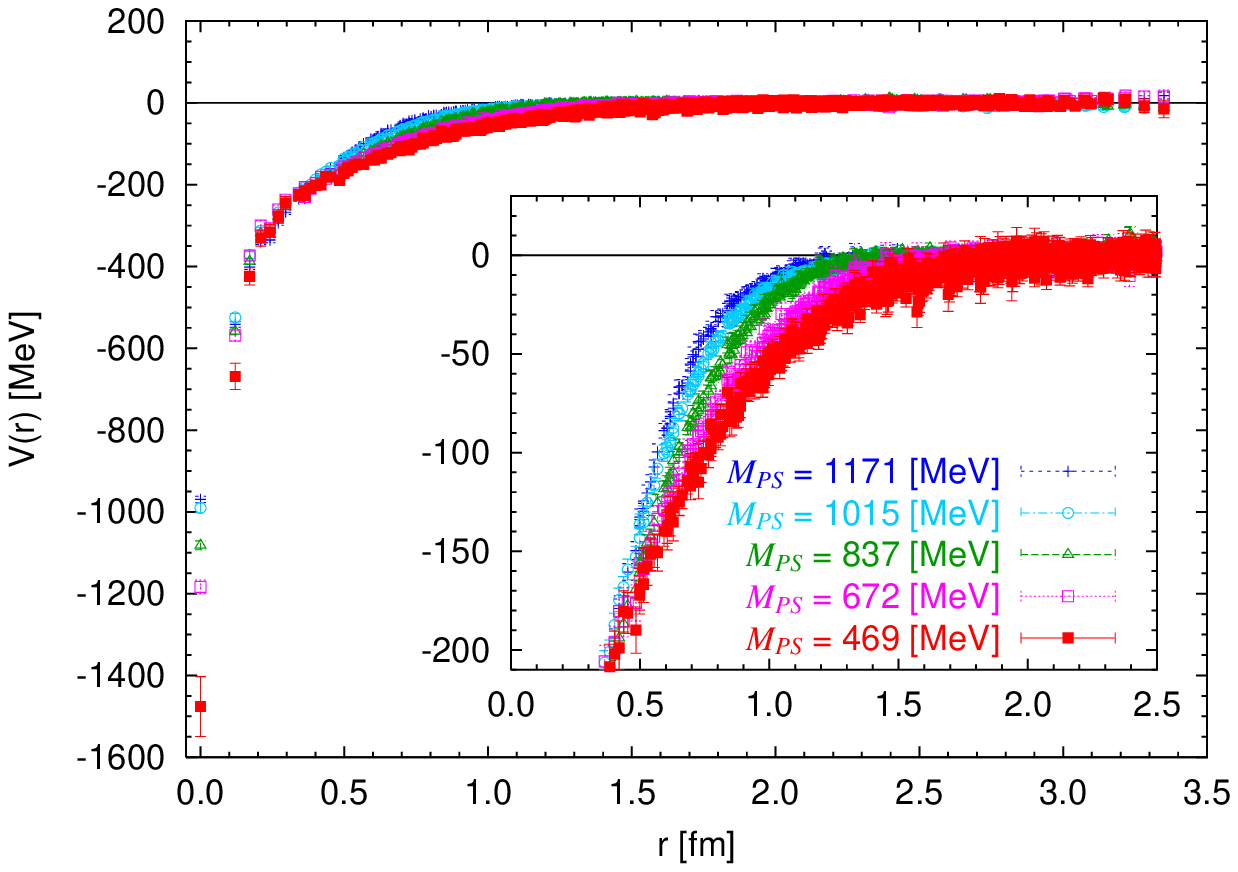}
\includegraphics[width=0.475\textwidth]{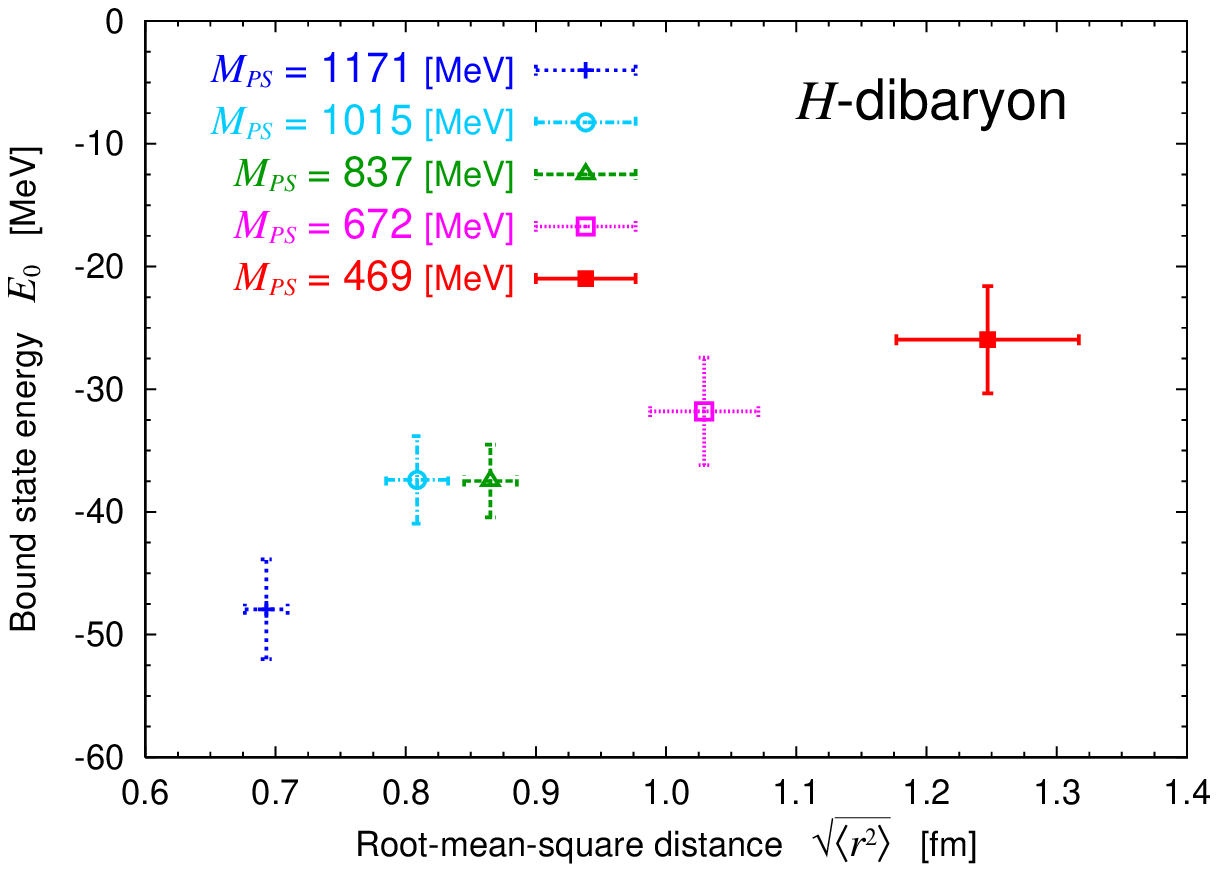}
\caption{ Left:  Potential of the flavor-singlet $B\!B$ channel extracted from QCD at five values of quark mass.
          Right: The ground state of the flavor-singlet $B\!B$ channel for each quark mass.}
\label{Fig1}
\end{figure}

 The left panel of Fig.\ref{Fig1} shows the flavor-singlet $B\!B$ potential
 extracted from our numerical simulation of full QCD at the five values of quark mass~\cite{Inoue:2011ai}.
 First of all, one sees that the potential is entirely attractive and has an attractive core.
 This feature does not seen in other $B\!B$ channels and hence is characteristic for this flavor singlet $B\!B$ channel.
 This lattice QCD result proves that Jaffe's and the other quark model prediction
 to the flavor-singlet $B\!B$ channel are essentially correct.
 
 By solving the Schr\"{o}dinger equation with these potential in the infinite volume,
 it turned out that there is one bound state for each quark mass~\cite{Inoue:2010es}.
 The right panel of Fig.\ref{Fig1} shows the energy and the root-mean-square (rms) distance of the bound states.
 These bound states correspond to the $H$-dibaryon predicted by Jaffe.
 Our lattice QCD calculation shows that a stable $H$-dibaryon certainly exists
 in the flavor $SU(3)$ limit world with the binding energy of 20--50 MeV for the present quark masses. 
 The rms distance $\sqrt{\langle r^2 \rangle}$ is a measure of size of the $H$-dibaryon,
 which can be compared to the point matter rms distance of the deuteron in nature: $3.8$ fm.
 From this comparison, one can get feeling of the $H$-dibaryon: It is much compact than deuteron.

\section{Effect of the flavor $SU(3)$ breaking on the H-dibaryon}

 In the real world, the flavor $SU(3)$ symmetry is broken and the $H$-dibaryon belongs to
 the $S=-2$, $I=0$, $B=2$ and $J^P=0^+$ sector, instead of the flavor singlet $B\!B$ channel.
 There exist three coupled $B\!B$ channels in this sector: $\Lambda\Lambda$, $N\Xi$ and $\Sigma\Sigma$.
 In order to study effect of the flavor $SU(3)$ breaking on the $H$-dibaryon,
 let us take into account the breaking in a phenomenological way.
 In practice, we combine the physical values of baryon masses
 and the lattice QCD $B\!B$ potentials extracted at a flavor $SU(3)$ limit
 with the pseudo-scalar meson mass $M_{PS}= 469$ MeV~\cite{Inoue:2011ai}.
 This calculation is based on the two assumptions:
 (i) that the major effect of the $SU(3)$ breaking comes from the baryon mass splittings, 
 (ii) and that the qualitative features of the hyperon interactions remain intact even with the $SU(3)$ breaking.

\begin{figure}[t]
 \includegraphics[width=0.475\textwidth]{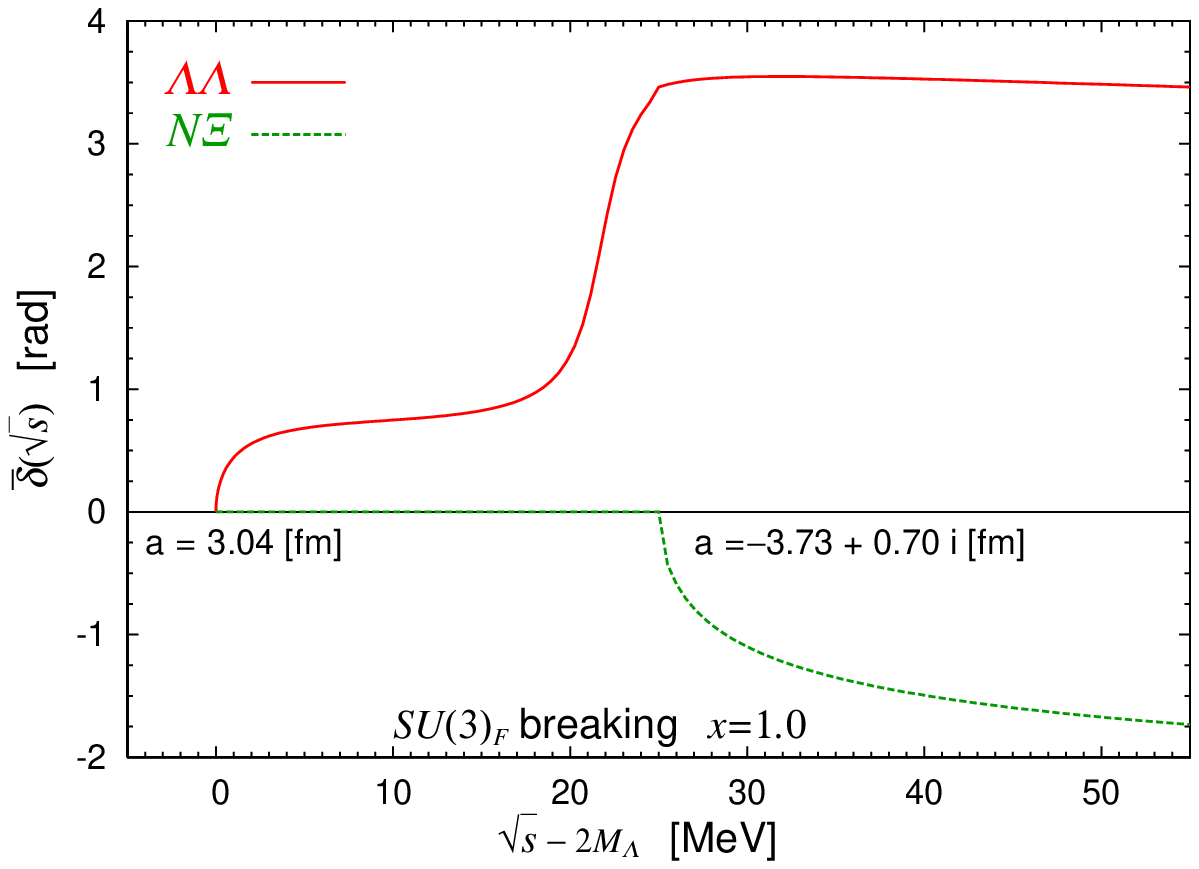}
 \includegraphics[width=0.475\textwidth]{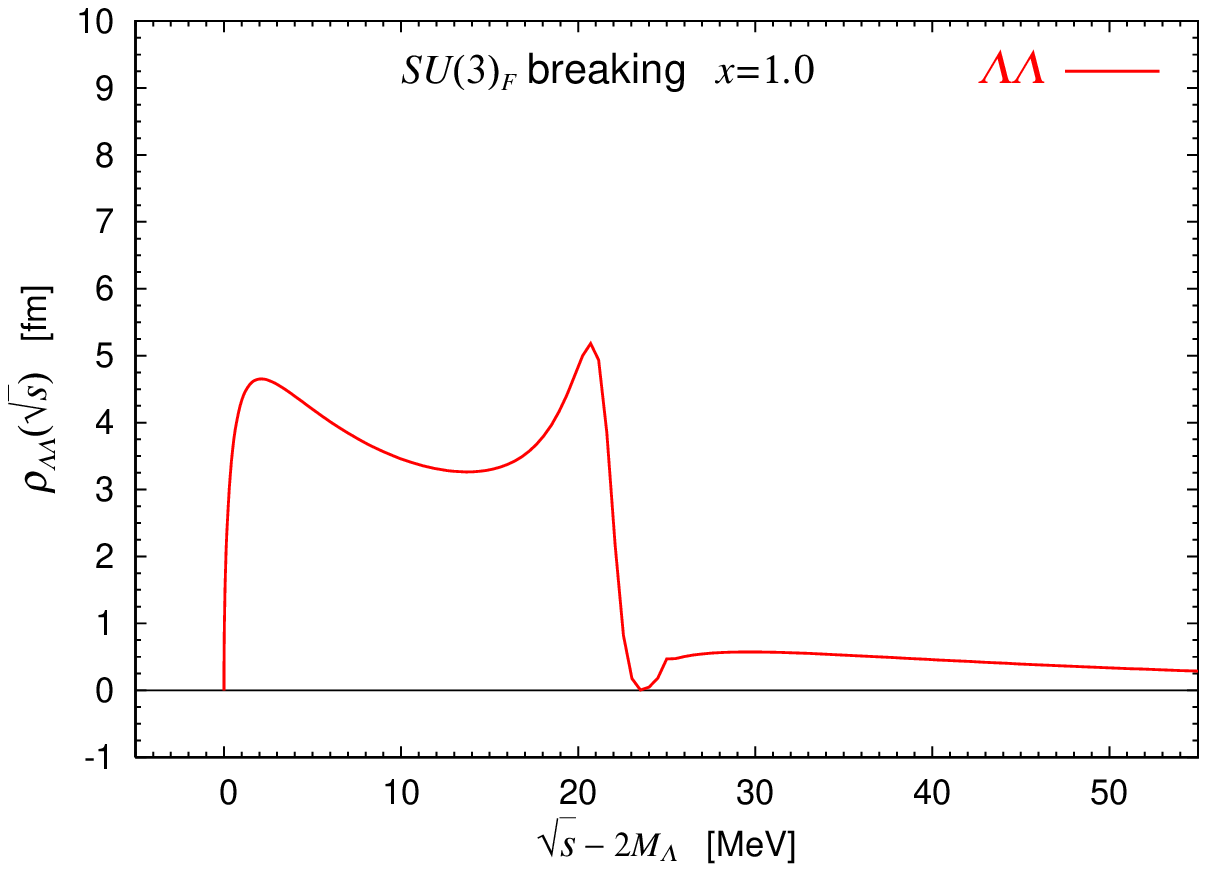}
\caption{
Left:  The bar-phase-shift of $\Lambda\Lambda$ and $N\Xi$. 
Right: The invariant-mass-spectrum of $\Lambda\Lambda$.
These are calculated with the lattice QCD potentials and the phenomenological baryon masses.
}
\label{fig:Fig2}
\end{figure}

 By solving the Lippmann-Schwinger equation for T-matrix,
\begin{equation}
 T^{\alpha\beta} = V^{\alpha\beta} + \sum_{\gamma}
 V^{\alpha\gamma} \, G^{(0)}_{\gamma} \, T^{\gamma\beta}, \quad 
 G^{(0)}_{\gamma} = \frac{1}{E - H^{(0)}_{\gamma} + i \epsilon}
\end{equation}
 we obtain S-matrix of the spin-singlet S-wave scattering in the coupled channels.
 The left panel of Fig.~\ref{fig:Fig2} shows the obtained bar-phase-shifts $\bar{\delta}_i$ 
 as a function of energy in the center of mass frame.
 We can clearly see a resonance in $\Lambda\Lambda$ bar-phase-shift at a little below the $N\Xi$ threshold.
 This resonance corresponds to the $H$-dibaryon. 
 In fact, when we increase a size of the breaking from zero to the physical one by hand, 
 we see that the bound state moves to upward and goes above the $\Lambda\Lambda$ threshold
 for the empirical flavor $SU(3)$ breaking. 
 If this phenomenological estimation is reasonable, 
 the $H$-dibaryon will be observed in nature as a resonance in $\Lambda\Lambda$ channel.
 Similar results are reported in refs.~\cite{Beane:2011xf,Shanahan:2011su,Haidenbauer:2011ah}. 
 However, to make a definite conclusion on this point from QCD,
 we need to carry out lattice QCD simulations at the physical point and analyses in the coupled channels. 
 Study toward this goal is in progress.

 The right panel of Fig.~\ref{fig:Fig2} shows the invariant-mass-spectrum
 of the $\Lambda\Lambda \to \Lambda\Lambda$ process calculated
 with the above phenomenological treatment of the flavor $SU(3)$ breaking and the S-wave dominance assumption.
 We can see a peak which corresponds to the $H$-dibaryon.
 This result demonstrates that there is a chance to confirm the existence of the resonant $H$-dibaryon in nature
 by experiments which count two $\Lambda$'s. 
 Actually, there is a report which says that an enhancement of the two $\Lambda$'s production has been observed
 at a little above the $\Lambda\Lambda$ thresholds in E224 and E522 experiments at KEK,
 although statistics is not sufficient for a definite conclusion~\cite{Yoon:2007aq}.
 New data with high statistics from an upgraded experiment at J-PARC~\cite{AhnImai}
 as well as the data from heavy-ion experiments at RHIC and LHC~\cite{Shah:2011en}
 will shed more lights on the nature of the $H$-dibaryon in near future.

\end{document}